\documentstyle[twoside,fleqn,espcrc2]{article}
\def\be{\begin{equation}}
\def\en{\end{equation}}
\def\bea{\begin{eqnarray}}
\def\ena{\end{eqnarray}}

\def\tr{{\rm tr}}

\title{Quantum gauge fixing and vortex dominance 
\vskip-3cm\hfill\small UNITU-THEP 11/99\vskip2.6cm
}
\author{K.~Langfeld\thanks{Presented by the first author 
at Lattice '99, Pisa, Italy.},
M.~Engelhardt, H.~Reinhardt, 
O.~Tennert, \address{Institut f\"ur Theoretische Physik, 
Universit\"at T\"ubingen, Auf der Morgenstelle 14, D-72076 T\"ubingen, 
Germany}}

\begin{document}
\begin{abstract}
We introduce quantum gauge fixing (QGF) as a new class of gauge 
fixings. While the maximal center gauge might not show 
vortex dominance, the confining properties of the vortices observed 
in past lattice calculations are argued to have been obtained 
in a gauge more akin to QGF than to the strict maximal center gauge. 

\end{abstract}
\maketitle
\section{Introduction}

In recent years, evidence has been accumulated that the mechanism of 
quark confinement may be understood in effective theories 
of monopoles~\cite{nam74,sch99} or $Z_N$ 
vortices~\cite{cgf}-\cite{la98}. 
After gauge fixing, these theories arise from projecting the full SU(N) 
onto a gauge theory with a reduced gauge symmetry. The observation 
that the reduced theories reveal the full string tension nurtures 
the conjecture that those degrees of freedom bear confinement. 
Here, we will employ center gauge fixing and center 
projection~\cite{cgf} for the reduction of SU(2) to $Z_2$ gauge theory 
which can be understood as a theory of vortices. We define {\it vortex 
dominance} if two criteria are met. Firstly, the string tension is 
preserved by projection. Secondly, the vortices survive the continuum 
limit. 

\section{ Center gauge fixing (standard) } 

Let $U_\mu (x)$ denote the link variable of SU(2) gauge theory and 
$\Omega (x)$ a gauge transformation matrix. Maximizing the functional 
\be 
S_{fix}[U] = \sum_ {\{x\} \mu } \frac{1}{2} \tr \left\{ U^\Omega  _\mu (x) 
\, \tau ^a \, U^{\Omega \, \dagger } _\mu (x) \, \tau ^a \right\} \; , 
\label{eq:1} 
\en 
with respect to $\Omega (x)$ yields the gauge transformation matrices 
which, when applied, cast a given configuration in the center gauge. 
{\it Center projection} is performed by the mapping of the 
link variable $U^\Omega (x)$ onto ${\pm 1 }$ (see e.g.~\cite{cgf}). 

In practical calculations, finding the absolute maximum of the 
functional $S_{fix}$ is a difficult task. The results which are 
presented in the next section were obtained by applying the algorithm 
presented in~\cite{cgf}, which resorts to iteration with 
over-relaxation. Once the iteration does not change the functional 
$S_{fix}$ any more up to a given precision, we performed a random 
gauge transformation on the actual link configuration and repeat the 
center gauge fixing. The procedure is repeated up to ten times. 
We finally choose the configuration $\Omega (x)$ which corresponds to the 
maximum value within the series of the ten fixing steps. 
This alleviates, but does not eliminate, the Gribov problem (see section~4).

\section{ Numerical results } 

{\it Zero temperature: } It turns out that the effective $Z_2$ gauge theory 
which was constructed with the ITOV-algorithm sketched above shows vortex 
dominance: the string tension is preserved~\cite{cgf,deb97}, and 
the linking number of the vortex lines with a Wilson loop 
(vortex density) meets with 
the expectations from a renormalization group analysis~\cite{la98}. 
Also the vortex interactions scale. These observations suggest that the 
vortices are physical objects surviving the continuum limit~\cite{la98}. 

\begin{figure}[htb]
\vspace{5.5cm}
\includegraphics{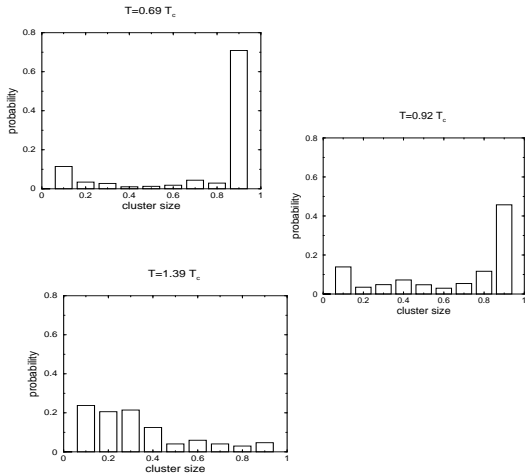}
\caption{ Distribution of the vortex cluster sizes; ''1'' is the maximum 
length possible in the lattice universe.}
\end{figure}
{\it Finite temperatures: } We find that below the deconfinement temperature 
$T_c$ vortex dominance persists~\cite{la99}. In particular, the effective 
$Z_2$ gauge theory correctly reproduces the critical temperature thus 
indicating that the essence of the deconfinement phase transition can be 
captured in the vortex picture. A thorough study of the density of 
vortices which are linked to spatially oriented Wilson loops shows that 
the vortex state at high temperatures is in agreement with the predictions 
of dimensional reduction~\cite{la99}. On the other hand, the density of
vortices linked with time-like Wilson loops only drops by a factor of 
three if the temperature is raised to twice $T_c$. 
Additional information is needed 
to understand the drop of the string tension to zero at $T_c$. 
We argue~\cite{la99,en99} that 
while the vortices are organized in a huge cluster at zero temperature, 
they stop percolating at $T_c$ and the huge cluster decays into many 
small size clusters. If in the later case the size of the Wilson loop 
exceeds the average cluster size, only vortices close to the 
circumference of the Wilson loop contribute yielding a perimeter law. 
In order to substantiate this idea, we measured the probability that 
a link of the vortex belongs to a cluster of given size. The result 
is depicted in figure 1. Further details can be found in~\cite{en99}. 

\section{ Gribov ambiguities } 

The variety of results in the previous section was obtained 
by implementing the center gauge (see (\ref{eq:1})) with the help of 
the ITOV algorithm sketched in section~2. It was 
recently pointed out that this procedure does not evade the Gribov 
problem~\cite{kov99}. In fact, it was observed that implementing the 
Landau gauge before applying the ITOV algorithm leads on average 
to a larger maximum value of $S_{fix}$ (\ref{eq:1}) than attained 
by the direct use of the ITOV algorithm. Moreover, the vortex 
state constructed from the gauge fixing via the Landau gauge detour 
does {\it not} show vortex dominance. These results cast doubt on the 
issue of vortex dominance in the strict {\it maximal } center gauge 
and call for a more accurate specification of the gauge in which the 
results outlined in section~3 were really obtained.

\section{ Quantum gauge fixing } 

For this specification, we here propose a new type of gauge fixing procedure 
which we will call {\it Quantum gauge fixing } (QGF). We will argue 
that the ITOV algorithm of section~2 already contains some of the 
characteristics of QGF rather than representing a numerically stable 
implementation of the maximal center gauge. Defining the 
$SU(N)/Z_N$ matrices 
\bea 
\zeta _{ab} (x) &:=& \frac{1}{2} \tr \left\{ \Omega (x) \, \tau ^a \, 
\Omega ^\dagger (x) \, \tau ^b \right\} \; , 
\label{eq:3}  \\ 
R^\mu _{ab} (x) &:=& \frac{1}{2} \tr \left\{ U_\mu (x) \, \tau ^a \, 
U_\mu ^\dagger (x) \, \tau ^b \right\} \; , 
\label{eq:4}  
\ena 
the gauge fixing functional $S_{fix}$ (\ref{eq:1}) can be cast into 
\be 
S_{fix} \; = \; \sum _{\{x\} \mu } \; \tr \; \zeta ^T (x+\mu ) \, 
R^\mu (x) \, \zeta (x) \; . 
\label{eq:5} 
\en 
Center gauge fixing corresponds to maximizing (\ref{eq:5}) with respect to 
the matrices $\zeta $ for a given lattice configuration, i.e., $R^\mu (x)$. 
From given matrices $\zeta (x)$, the gauge transformation $\Omega (x)$ 
can be constructed up to a center gauge transformation, 
\be 
f^Z_{SU}: \; SU(N)/Z_N \rightarrow SU(N): \; \Omega = f^Z_{SU}(\zeta ) 
\; . 
\label{eq:6} 
\en 
This ambiguity reflects the familiar fact that the maximal center gauge 
condition leaves a residual $Z_N$ gauge group unfixed. 

We define the QGF as follows: by means of a functional integral over gauge 
transformations $\zeta $, we construct the matrix 
\be 
\omega [U] \; := \;  \frac{1}{N} \int {\cal D} \zeta \; f^Z_{SU}(\zeta ) \; 
\exp \left\{ \beta _f \, S_{fix} \right\} \; , 
\label{eq:7} 
\en 
which in general is not an element of the gauge group. The gauge 
transformation 
$\Omega (x)$ which brings a given field configuration $U_\mu (x)$ 
into the quantum gauge is then defined by the SU(N) element ''closest'' 
to $\omega [U]$, i.e., 
\be 
\vert \vert \, \omega [U](x) - \Omega (x) \, 
\vert \vert ^2 \rightarrow \hbox{min} \; , \; \; \; \forall \, x \; , 
\label{eq:8} 
\en 
where $\vert \vert A \vert \vert ^2 := \tr A A^\dagger $. 
It can be shown that the QGF (\ref{eq:7}-\ref{eq:8}) is free of Gribov 
ambiguities. In the case of the center gauge fixing (\ref{eq:5}), the 
functional integral in (\ref{eq:7}) can be viewed as a partition function 
of the matrices $\zeta $ each interacting with its nearest neighbors 
via the metric $R^\mu (x)$. This quantum theory of matrices $\zeta $ is 
therefore interpreted as generalized spin glass. 
Spin glass systems are known for a complex phase structure. 
A numerical calculation of the expectation value (\ref{eq:7}) 
is therefore tedious, and we recover the Gribov problem in practical 
applications.

\begin{figure}[htb]
\vspace{4.5cm}
\includegraphics{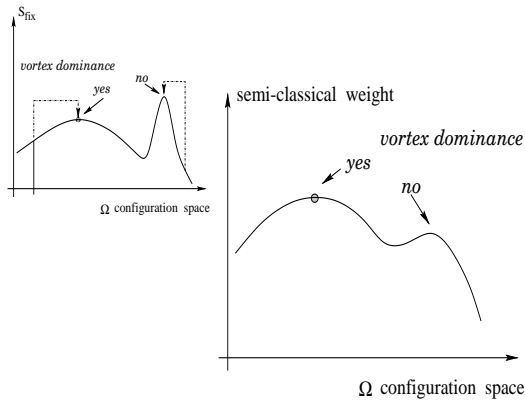}
\caption{ The illustration of $S_{fix}$ and the semi-classical weight 
   (\protect\ref{eq:9}) as functional of $\zeta (x)$. 
}
\end{figure}
Nevertheless, QGF is helpful to put the results outlined in section~3 
in the proper context. For this purpose, we consider large values 
of the gauge parameter $\beta _f $ in (\ref{eq:7}). In this case, the 
configuration $\zeta (x)$ which corresponds to a local maximum of 
$S_{fix}$ contributes to $\omega $ with the (semi-classical) weight 
\be 
\propto \, \exp \left\{ \beta _f S_{fix} \left[\zeta \right] \right\} 
\, / \, \left[ S^{\prime \prime }_{fix} \left[ \zeta \right] 
\right] ^{1/2} \; , 
\label{eq:9} 
\en 
where $S^{\prime \prime }_{fix}$ is the functional determinant 
of the second (functional) derivative of $S_{fix}$ with respect to 
$\zeta (x)$. Eq.(\ref{eq:9}) implies that the contributions of 
maxima with large curvature are suppressed. The situation is illustrated 
in figure~2. 
Since the ITOV algorithm of section~2 involves several steps of iteration 
which start with a random choice of $\zeta $, it probes the 
volume of the region of attraction corresponding to a particular maximum. 
The ITOV algorithm therefore 
effectively contains a similar entropy factor as the 
one implied by (\ref{eq:7}). We therefore argue that the results shown in 
section~3 refer to a gauge which is more akin (but not completely 
identical) to QGF than to the strict maximal center gauge. 

{\tt Supported in part by DFG-RE 856/4-1, DFG-En 415/1-1. }


\begin{thebibliography}{9}
\bibitem{nam74}{ Y.~Nambu, Phys. Rev. {\bf D 10} (1974) 4262; \\
   G.~'t~Hooft, in: {\em High Energy Physics}, ed. A.~Zichichi, 
   Bologna, 1975; \\
   S.~Mandelstam, Phys. Rep. {\bf B 23} (1976) 245. } 
\bibitem{sch99}{ K.~Schilling, G.S.~Bali and C.~Schlichter,
   Nucl. Phys. Proc. Suppl. {\bf 73} (1999) 638. \\ 
   A.~Di Giacomo, B.~Lucini, L.~Montesi and G.~Paffuti,
   hep-lat/9906024. } 
\bibitem{cgf}{ L.~Del Debbio, M.~Faber, J.~Giedt, J.~Greensite and S.~Olejnik,
   Phys. Rev. {\bf D58} (1998) 094501; \\ 
   L.~Del Debbio, M.~Faber, J.~Greensite and S.~Olejnik,
   Nucl. Phys. Proc. Suppl. {\bf 53} (1997) 141. } 
\bibitem{tho78}{ G.~'t~Hooft, Nucl. Phys. {\bf B138} (1978) 1; \\
   G.~Mack, in: {\em Recent Developments in Gauge Theories},
   eds. G.~'t~Hooft et al (Plenum, New York, 1980). } 
\bibitem{deb97}{ L.~Del Debbio, M.~Faber, J.~Greensite and S.~Olejnik,
   Phys. Rev. {\bf D55} (1997) 2298. } 
\bibitem{la98}{ K.~Langfeld, H.~Reinhardt and O.~Tennert,
   Phys. Lett. {\bf B419} (1998) 317; \\ 
   M.~Engelhardt, K.~Langfeld, H.~Reinhardt and O.~Tennert,
   Phys. Lett. {\bf B431} (1998) 141. } 
\bibitem{la99}{ K.~Langfeld, O.~Tennert, M.~Engelhardt and H.~Reinhardt,
   Phys. Lett. {\bf B452} (1999) 301. } 
\bibitem{en99}{ M.~Engelhardt, K.~Langfeld, H.~Reinhardt and O.~Tennert,
   hep-lat/9904004. } 
\bibitem{kov99}{ T.~G.~Kovacs, E.~T.~Tomboulis, hep-lat/9905029. } 

\end{thebibliography}
\end{document}